\documentclass[preprint]{vgtc}               

\graphicspath{{figures/}{pictures/}{images/}{./}} 

\usepackage{times}                     

\usepackage{tabu}                      
\usepackage{booktabs}                  
\usepackage{lipsum}                    
\usepackage{mwe}                       
\usepackage{mathptmx}                  
\usepackage{soul}

\usepackage{xcolor}
\usepackage{enumitem}
\usepackage{mathptmx}                  
\usepackage{amsmath}                   
\setlist{
  itemsep=2pt,
  topsep=4pt,
  parsep=0pt,
  partopsep=0pt
}

\def\etal{{\textit{et~al.}}}
\onlineid{0}

\vgtccategory{Research}

\vgtcinsertpkg

\title{KuroshioVis: Hypothesis-Driven Discovery of \\Mesoscale Air-Sea Interaction through Coarse-to-Fine Exploration
}

\author{Yuanzhi Chen\\ %
        \scriptsize Shandong University %
\and Yiwei Lin\\ %
     \scriptsize Shandong University %
\and Xi Duan\\ %
     \scriptsize Shandong University %
\and Qingqing He\\ %
     \scriptsize Shandong University %
\and Ruiliang Xu\\ %
     \scriptsize Shandong University %
\and Yucheng Wang\\ %
     \scriptsize Laoshan Laboratory %
\and Qiong Zeng\thanks{qiong.zn@sdu.edu.cn}\\ %
     \scriptsize Shandong University %
     }

\teaser{
  \centering
  \includegraphics[width=\linewidth]{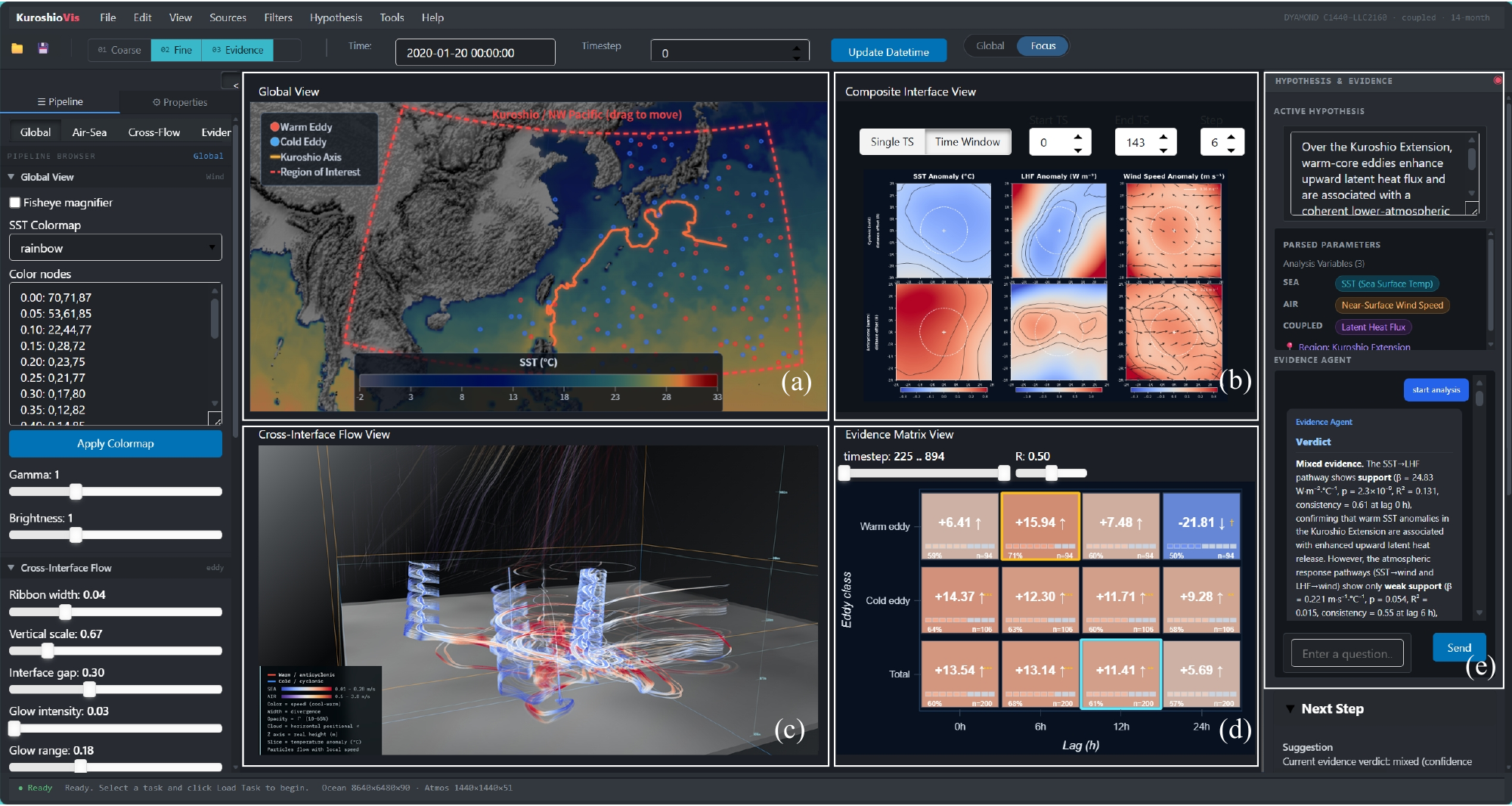}
  \caption{\emph{KuroshioVis} supports the hypothesis-driven visual analytics of ocean-atmosphere coupling through a coarse-to-fine workflow. The system integrates coordinated views for global event localization (a), multivariate composite visualization of the air-sea interface (b), cross-interface flow visualization (c), and statistical evidence analysis (d), enabling users to identify candidate coupling events, examine their physical structures, quantify statistical support, and interactively refine scientific hypotheses through AI-assisted reasoning (e).}
  \label{fig:teaser}
}

\abstract{
Mesoscale air-sea interaction occurs throughout strong sea surface temperature fronts such as the the Kuroshio Extension, but its strength is highly non-uniform and can be resolved only in kilometre-scale coupled simulations. 
The response of any single eddy is weak relative to synoptic variability, so evidence supporting a hypothesis of systematic air-sea coupling must be accumulated across many eddies rather than inferred from individual cases. We present \textit{KuroshioVis}, a hypothesis-driven, coarse-to-fine visual analytics workflow. The coarse stage extracts and aligns mesoscale eddies into composite representations of their air-sea responses, and provides a \emph{global view} for exploration. The fine stage investigates these responses through three coordinated views: the \emph{composite interface view}, the \emph{cross-interface flow view}, and the \emph{evidence matrix View}. These coordinated views support air-sea coupling analysis from surface patterns to cross-interface dynamics and quantitative statistical evidence. We demonstrate the workflow using the coupled C1440-LLC2160 simulation of the 2026 SciVis Contest.
}


\begin{document}

\maketitle

\section{Introduction}
Mesoscale air-sea interaction over strong sea surface temperature (SST) frontal regions such as the Kuroshio Extension plays an important role in modulating turbulent heat exchange, atmospheric boundary-layer adjustment, and regional climate variability~\cite{Small2008,Byrne2016,Seo2023}. 
Recent kilometre-scale coupled ocean-atmosphere simulations have resolved coherent atmospheric responses to mesoscale oceanic features, including warm- and cold-core eddies~\cite{Seo2023}. However, these responses are often weak relative to synoptic variability, spatially heterogeneous, and temporally intermittent~\cite{Strobach2022}. Individual events therefore provide limited evidence on their own, while the relevant signatures are distributed across oceanic and atmospheric variables, spatial structures, and temporal lags. The challenge is thus to organize responses across multiple events into coherent evidence for testing a scientific hypothesis.

Visualization has become an important means of exploring large-scale oceanographic and atmospheric simulations~\cite{Afzal2019}. Existing systems support ocean forecast analysis~\cite{Hollt2014}, in situ feature extraction from high-resolution simulations~\cite{Woodring2016}, Lagrangian eddy boundary identification~\cite{Friederici2018}, narrative exploration of eddy evolution~\cite{Han2022}, and scalable visual analysis of large ocean datasets~\cite{Jain2025}. 
These approaches have substantially advanced the exploration of ocean variables, mesoscale structures, and their temporal evolution. However, most focus on individual variables, events, or processes within a single physical domain.
Air-sea interaction instead links oceanic variability to atmospheric adjustment through exchanges of heat, momentum, and mass~\cite{Seo2023}, with relevant evidence distributed across both domains, multiple events, spatial structures, and temporal lags. Its analysis therefore requires examining both oceanic and atmospheric processes and evaluating the statistical support for their coupling under an explicit hypothesis.

We present \textit{KuroshioVis}, a hypothesis-driven, coarse-to-fine visualization workflow that organizes analysis around scientific hypotheses. The coarse stage (\cref{subsec:coarse}) extracts and aligns mesoscale eddies, aggregating them into an ensemble representation; and provides a \emph{global view} (\cref{fig:teaser}a) to display the extracted eddies and the jet axis.
The fine stage (\cref{subsec:fine}) then characterizes this ensemble through three coordinated views: the \emph{composite interface view}, the \emph{cross-interface flow view}, and the \emph{evidence matrix view}. An AI agent (\cref{fig:teaser}e) supports the workflow by translating the hypothesis into analysis criteria that select an event subset and retrieve the corresponding statistical evidence, and by reporting which sub-claims are supported by the accumulated evidence. 
Although instantiated for the Kuroshio Extension, the workflow can be adapted to other frontal regions in which recurring mesoscale events can be extracted, aligned, and evaluated under explicit hypotheses.

Specifically, the \emph{composite interface view}  (\cref{fig:teaser}b) provides an ensemble-based multivariate composite visualization that integrates multiple coupled surface variables within a common eddy-centric reference frame while preserving both the ensemble mean and response consistency.
The \emph{cross-interface flow view} (\cref{fig:teaser}c) presents an eddy-ensemble flow visualization that overlays oceanic and atmospheric stream ribbons derived from aligned eddy instances, highlighting flow patterns that remain consistent across the ensemble on both sides of the air-sea interface.
The \emph{evidence matrix view} (\cref{fig:teaser}d) organizes statistical evidence across eddy categories and atmospheric lags, allowing analysts to examine the regression statistics and corresponding scatterplot for each matrix cell.

We evaluate the workflow on the coupled C1440-LLC2160 simulation provided for Task 2 of the 2026 IEEE SciVis Contest. A case study traces how a scientific hypothesis is examined through \emph{KuroshioVis}, constructing a diagnostically relevant eddy ensemble in the coarse stage to inspecting complementary field, flow, and statistical evidence in the fine stage. 
We further conduct a demonstration-based evaluation with two physical oceanographers, who assess the scientific plausibility, interpretability, and practical value of the workflow.
\begin{figure*}[htbp]
\centering
\includegraphics[width=1\linewidth]{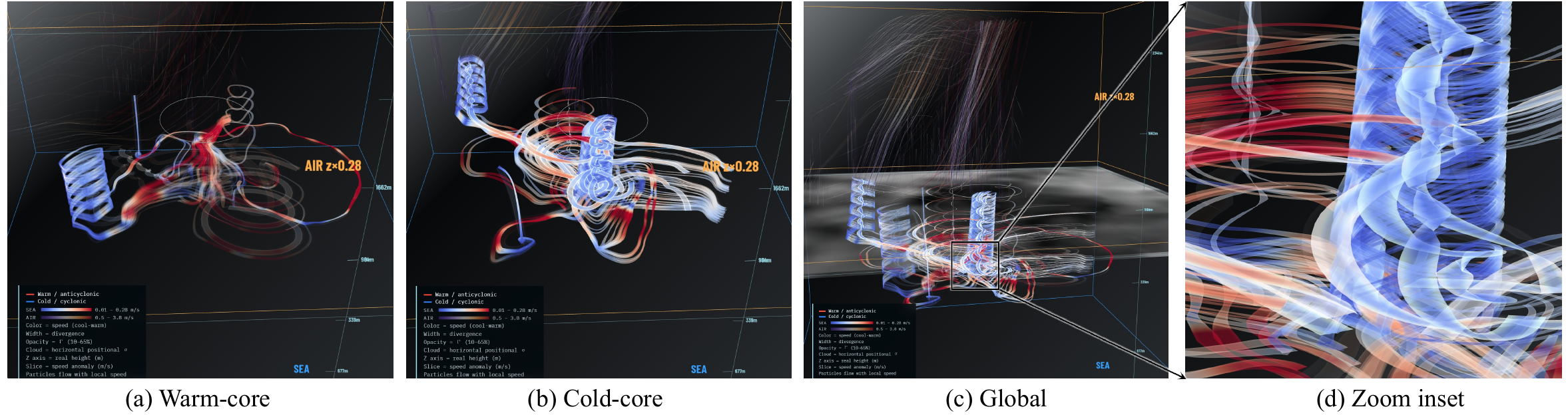}

\caption{\textbf{Cross-interface flow view} comparing warm- and cold-core eddy ensembles. (a) and (b) show the warm- and cold-core ensembles in the same eddy-centered, wind-aligned frame using identical seeds, scaling, rendering parameters, and camera settings. Oceanic and lower-atmospheric trajectories are rendered as ribbons on opposite sides of the air-sea interface, with color encoding speed. (c) presents the global configuration, showing the relative spatial organization of the oceanic and atmospheric flows together with the contextual SST slice. (d) provides a zoomed inset of the oceanic circulation, highlighting its compact rotational structure and local variations in ribbon width and twist.}
\vspace*{-1mm}
\label{fig:overview}
\end{figure*}

\section{Approach}
\subsection{Coarse Stage}
\label{subsec:coarse}
The coarse stage extracts the jet axis (\cref{subsubsec:jet}), detects and aligns mesoscale eddies (\cref{subsubsec:eddy}), aggregates the aligned eddies into an ensemble representation (\cref{subsubsec:ensemblerep}) for the fine stage (\cref{subsec:fine}), and presents these elements in the \emph{global view} (\cref{subsubsec:global}).

\subsubsection{Jet Axis Extraction}
\label{subsubsec:jet}
We extract the Kuroshio Extension jet axis at every time step using a velocity-guided algorithm that selects the sea surface height (SSH) contour best aligned with surface-current speed maxima. The extracted axis serves as a spatial reference in the \emph{global view}, helping analysts locate eddies relative to the main jet. Note that the jet axis does not enter the composite or statistical analysis that follow.
 
\subsubsection{Eddy Detection and Alignment}
\label{subsubsec:eddy}
We detect mesoscale eddies from SSH using the closed-contour method of Chelton~\etal~\cite{Chelton2011} and the implementation of Oliver~\cite{Oliver2015}. We identify each eddy as the outermost closed SSH contour enclosing a single extremum, record its center, equivalent radius, amplitude, and shape error, and classify it as warm-core (anticyclonic) or cold-core (cyclonic) according to its polarity. We then link detections across time steps into trajectories, from which we derive eddy lifetime and propagation. 
To support further ensemble analysis, we transform each eddy into a common eddy-centered, wind-aligned frame by translating its center to the origin, normalizing horizontal coordinates by the equivalent radius, and rotating the positive $x$-axis toward the mean lower-atmospheric wind computed over an annulus from $1.5R$ to $2R$.

\subsubsection{Ensemble Representation}
\label{subsubsec:ensemblerep}
Given the aligned eddies, the coarse stage constructs the representation passed to the fine stage:
\begin{equation}
\mathbf{R}(\mathcal{C})
=
\left\{
\bar{\mathbf{X}},
\Gamma,
S
\right\},
\end{equation}
where $\bar{\mathbf{X}}$ denotes the composite response, $\Gamma$ the consistency field, and $S$ the statistical-significance field. For each eddy $e_i$, let
$\mathbf{X}_i=(\mathrm{SST}'_i,Q'_{E,i},\mathbf{U}'_i)$
denote its surface and near-surface anomaly fields, computed relative to the corresponding mean over an annulus spanning two to three eddy radii. Here, $\mathrm{SST}'_i$ is the sea surface temperature anomaly, $Q'_{E,i}$ is the upward latent heat flux anomaly, and $\mathbf{U}'_i$ is the near-surface horizontal wind anomaly.

We transform $\mathbf{X}_i$ into the common eddy-centered, wind-aligned frame and resample it onto a normalized grid with coordinates $\mathbf{x}$. In this frame, the eddy center is placed at the origin, horizontal distances are normalized by the equivalent eddy radius, and the positive $x$-axis is aligned with the local mean wind direction. This transformation yields the aligned field $\hat{\mathbf{X}}_i(\mathbf{x})$, enabling point-wise aggregation and comparison across eddies:

\begin{equation}
\bar{\mathbf{X}}(\mathbf{x})
=
\frac{1}{|\mathcal{C}|}
\sum_{e_i\in\mathcal{C}}
\hat{\mathbf{X}}_i(\mathbf{x}).
\end{equation}
We further evaluate how consistently individual eddies exhibit the expected response for warm-core and cold-core groups separately:
\begin{equation}
\Gamma(\mathbf{x})
=
\frac{1}{|\mathcal{C}|}
\sum_{e_i\in\mathcal{C}}
\mathbf{1}
\!\left(
\operatorname{sign}
(\hat{\mathbf{X}}_i(\mathbf{x}))
=
\sigma_k
\right),
\end{equation}
where $\sigma_k$ is the expected response sign for the group. $\Gamma$ reports the proportion of eddies whose local response agrees with the expected sign at each location. The consistency field requires an expected sign $\sigma_k$, which warm-core and cold-core eddies provide directly, though the construction extends to any trackable feature with a specifiable expected response.
Significance $S$ is estimated by a one-sample $t$-test across the aligned samples.

 \subsubsection{Global View}
\label{subsubsec:global}
The \emph{global view} places the Kuroshio jet axis and detected warm- and cold-core eddies on an interactive globe, providing geographic context for the user-selected spatial and temporal range. The selected region is outlined directly on the globe, while the eddies satisfying the current filters are overlaid relative to the jet axis. SST is mapped over the ocean to reveal the large-scale thermal structure and SST fronts surrounding the detected eddies. We apply precomputed elevation shading to the land surface to emphasize terrain relief and visually distinguish land from ocean.

\subsection{Fine Stage}
\label{subsec:fine}
The fine stage examines the ensemble representation produced in the coarse stage (\cref{subsec:coarse}) through the following coordinated views.

\subsubsection{Composite Interface View}
The \emph{composite interface view} represents the aligned ensemble as eddy-centered composite maps of three surface variables: SST, latent heat flux (LHF), and near-surface wind anomalies. The view encodes the wind-aligned ensemble mean of SST and LHF anomalies using the \emph{coolwarm} colormap and overlays the composite wind anomaly as a vector field in the same normalized coordinates. Contours indicate the spatial structure of the composite responses, while the dashed circle marks the normalized eddy boundary. Warm-core and cold-core composites are displayed using shared coordinates, colormaps, and visual encodings, enabling direct comparison of their surface responses across the three variables. 

\subsection{Cross-Interface Flow View}
The \emph{cross-interface flow view} constructs and visualizes oceanic and lower-atmospheric trajectories in a shared eddy-centered, wind-aligned frame, enabling direct comparison of flow organization on both sides of the air-sea interface and across eddy polarities.

\subsubsection{Eddy-Centered Flow}
Using the normalized reference frame defined above, we place identical seeds for all eddy instances and map them to their corresponding physical coordinates. For the ocean, the mean velocity in an annulus between $1.6R$ and $3R$ is treated as the local background advection and subtracted before trajectory integration~\cite{Gaube15}. This isolates the eddy-relative circulation from the large-scale flow that translates or deforms the eddy~\cite{Chelton11}. We then trace oceanic trajectories through the three-dimensional velocity field and integrate atmospheric trajectories independently at the lowest available atmospheric level using corresponding horizontal seeds. We leave the unsampled region between this level and the sea surface empty rather than interpolating it.

We provide two complementary representations of oceanic circulation. In the ring representation, we sample the flow at multiple radii and depths and use vertical velocity to control the displayed helical pitch. In the layered representation, we integrate streamlines using a second-order Runge--Kutta method~\cite{Butcher16} and retain only those whose rotation is sufficiently coherent and consistent with the detected eddy polarity. We then transform the valid oceanic and atmospheric trajectories back into the normalized reference frame and organize them according to the polarity-specific eddy ensembles defined above, excluding instances that do not satisfy the prescribed trajectory-quality criteria.

\subsubsection{Visual Design}
We design the \emph{cross-interface flow view} to emphasize the compact rotational structure of the oceanic eddy while preserving its relationship to the more spatially extended atmospheric flow. To keep both domains legible within a shared three-dimensional space, we visually separate them at the air-sea interface and apply independent vertical scaling. We render trajectories as camera-facing ribbons with interactively adjustable base widths, while variations in ribbon width and twist reveal changes along individual flow paths.

To establish a clear visual hierarchy, we encode oceanic and atmospheric speeds using separate colormaps and emphasize the oceanic circulation with more salient colors. A vertical slicing plane reveals the SST distribution at different depths using a muted colormap, providing thermal context without competing with the trajectory structures. Animated particles further indicate flow direction and relative speed, making the rotational organization of the oceanic flow and its relation to the atmospheric circulation easier to interpret. We use identical normalization, rendering parameters, and camera settings for warm- and cold-core eddies, enabling direct comparison of their circulation patterns.

\cref{fig:overview} compares the warm- and cold-core eddy ensembles. (a) and (b) show the two ensembles separately, while (c) overlays them to support direct comparison of their oceanic and atmospheric trajectories. (d) provides a close-up view of the oceanic circulation, revealing details that are difficult to perceive at the global scale, including ribbon width variation, twist, and the distribution of positional uncertainty.

\subsection{Evidence Matrix View}
\begin{figure}[htbp]
\centering
\includegraphics[width=0.6\linewidth]{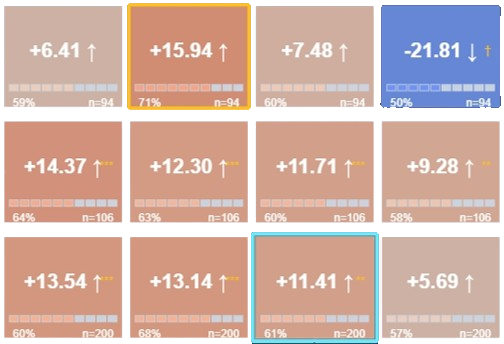}
\caption{\textbf{Evidence Matrix View}}
\vspace*{-1mm}
\label{fig:evidence_matrix}
\end{figure}

The \emph{evidence matrix view} organizes statistical evidence across eddy categories and atmospheric response lags. Its rows represent the warm-core, cold-core, and pooled eddy ensembles, while its columns represent atmospheric lags relative to the oceanic signal. For each eddy category and lag combination, we fit a linear regression, $y=\beta x+\varepsilon$, between the oceanic and atmospheric variables specified by the active hypothesis, such as SST anomaly and latent-heat-flux anomaly (\cref{fig:teaser}d).
Each cell encodes the signed regression slope $\beta$ through color and displays its value, statistical significance, and response consistency, allowing analysts to compare the strength and reliability of the inferred relationships across eddy categories and lags. Hovering over a cell reveals the corresponding scatterplot and detailed regression statistics, see \cref{fig:evidence_matrix}.

\subsection{AI Agent}
\label{subsec:agent}

The AI agent supports hypothesis evaluation through a three-step process. It first structures the scientific hypothesis into testable sub-claims and expected variable relationships, such as SST to LHF. It then collects relevant evidence from the structured outputs underlying the four views using predefined evidence-collection rules, which specify the view-derived sources to query, the files or runtime states to read, the metrics to extract, and the registered Evidence Matrix datasets to include in the evidence payload, rather than interpreting the rendered images directly.
Based on the collected evidence, the language model evaluates the sub-claims and summarizes the key evidence according to the expected variable relationships. It provides next-step guidance that aggregates evidence for longer-window analyses across event, monthly, seasonal, and multi-year time scales.
\begin{figure}[htbp]
\centering
\includegraphics[width=1\linewidth]{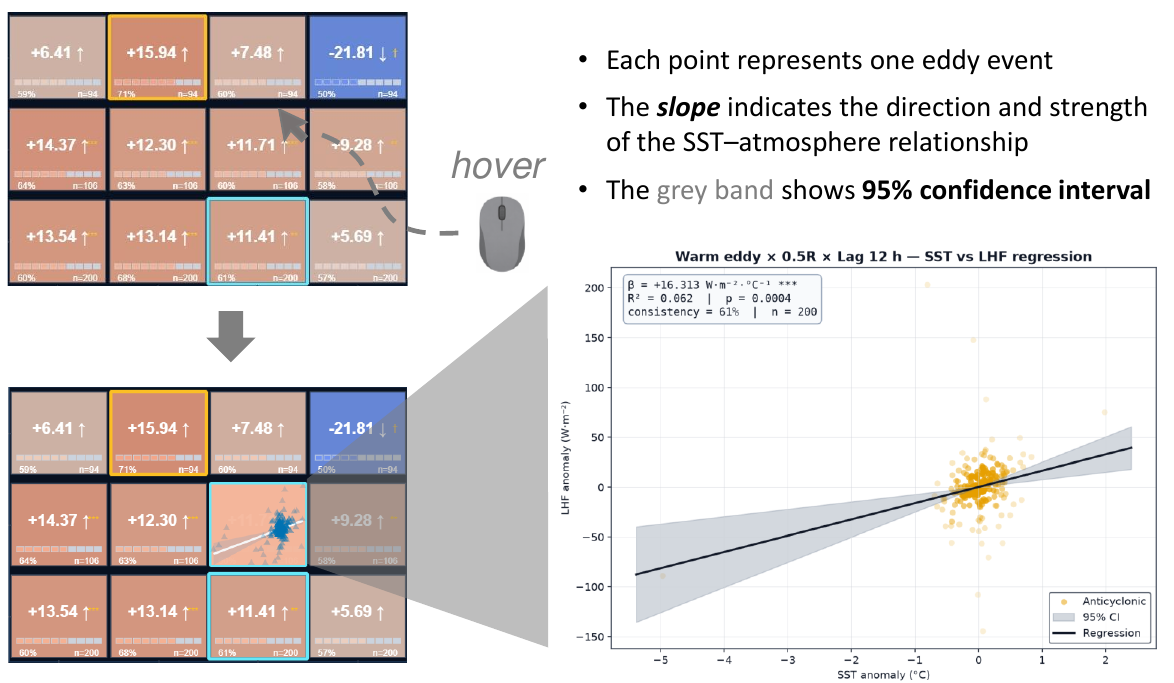}
\caption{\textbf{Hover interaction in the Evidence Matrix View}.
Each cell summarizes a regression relationship for a specific eddy condition and atmospheric lag. Hovering over a cell reveals the underlying event-level scatterplot, fitted regression line, 95\% confidence interval, and detailed statistics, allowing analysts to inspect the evidence behind the matrix summary.}
\vspace*{-1mm}
\label{fig:overview}
\end{figure}

\section{Results}
\subsection{Implementation Details}
We implement the system in Python. We first download the required subsets of the DYAMOND C1440 and ECCO LLC2160 dataset~\cite{Menemenlis2026} through OpenVisus~\cite{Pascucci2012Visus, Panta2024, OpenVisus}, and then perform all preprocessing and analysis locally. To support joint analysis of the two datasets, we map their cubed-sphere and MITgcm LLC grids onto a common grid using \texttt{scipy.interpolate.griddata}. We develop the interactive views with Panel and WebGL, and use deepseek-v4-flash~\cite{deepseekai2026deepseekv4} to support the AI agent. All experiments were conducted on an i7-13700H/RTX 4060 laptop with 16 GB memory.

\subsection{Case Study}
We illustrate the workflow using the following hypothesis over the Kuroshio Extension: \emph{warm-core eddies enhance upward latent heat flux and are associated with a coherent lower-atmospheric wind response, with the coupling strongest at short atmospheric lags}. The AI agent first structures the hypothesis into expected relationships among SST, LHF, and near-surface wind speed. Using the shared analysis controls, the analyst then specifies the Kuroshio region ($100^\circ$E-$150^\circ$E, $15^\circ$N-$40^\circ$N), the temporal range from January 29 to February 29, 2020, and the relevant variables.

The analysis begins in the \emph{global view}, where the analyst obtains a set of accepted eddy instances (\cref{fig:teaser}a).
The analyst then examines the ensemble-level surface responses in the \emph{composite interface view} (\cref{fig:teaser}b). The warm-core composite exhibits a positive SST anomaly, whereas the cold-core composite shows the opposite thermal signature. The corresponding LHF composites reveal polarity-dependent heat-flux responses around the normalized eddy center, while the overlaid wind vectors show how the near-surface atmospheric flow differs between the two eddy categories. These observations provide visual support for the SST-LHF component of the hypothesis and indicate how the atmospheric response is spatially organized relative to the eddy.

The analyst next uses the \emph{cross-interface flow view} to examine the relationship between the compact oceanic circulation and the more spatially extended lower-atmospheric flow, assessing whether the atmospheric trajectories exhibit a consistent organization relative to the underlying eddy circulation.
Furthermore, the analyst explores the \emph{evidence matrix view} to evaluate the expected variable relationships for warm-core, cold-core, and pooled eddy ensembles at atmospheric lags of 0, 6, 12, and 24 hours. For the SST-LHF relationship, the warm-core ensemble exhibits the strongest positive regression slopes at 6 hours, with $\beta=15.94$, respectively. The relationship weakens at 12 hours and becomes negative at 24 hours. This pattern supports an enhanced short-lag heat-flux response over warm-core eddies but does not support a persistent response at longer lags.

\subsection{Expert Study}
We evaluated the system with two domain experts in oceanography: \textbf{E1} had 10 years of experience in small- and mesoscale ocean processes; while \textbf{E2}, one of the co-authors, had 15 years of experience in ocean ecological dynamics in the northwestern Pacific and Kuroshio region. 
We presented a predefined air-sea coupling hypothesis and demonstrated the complete workflow using the same dataset and analysis settings to both experts. Rather than operating the system directly, both experts assessed the physical plausibility of the results, the coherence of the analysis process, and its potential value for scientific practice.

The experts first assessed the physical interpretability of the visual representations. Particularly, E1 noted that the results in the cross-interface flow view ``\emph{clearly showed the rotational structure of the eddies and its correspondence with sea surface temperature}''. E1 further explained that, \emph{although vertical motion is much weaker than horizontal flow, it represents processes such as upwelling and nutrient transport and is therefore conceptually very important}. 
The experts also assessed the overall analysis workflow. E2 described the overall framework from regional analysis to statistics, visualization, and numerical conclusions as ``\emph{logically clear}''. 
When asked whether the coordinated visualization of multiple views imposed excessive cognitive demand, E2 responded that ``\emph{it was not too much at all}'' and explaining that ``\emph{oceanographers commonly compare many plots when verifying relationships across diagnostics}''. Regarding the AI agent, E2 considered ``\emph{it potentially useful when its conclusions were organized around the underlying data rather than generated as generic text}'', and suggested incorporating domain knowledge and scientific literature to support a more specialized agent in future work.

\section{Conclusion}
We present \textit{KuroshioVis}, a hypothesis-driven, coarse-to-fine visual analytics system for investigating mesoscale air-sea interaction in the Kuroshio Extension. \textit{KuroshioVis} aligns eddy events into ensemble representations and coordinates multiple views to support analysis. In contrast to conventional visualization systems that primarily display individual variables or events, \textit{KuroshioVis} organizes both computation and visual exploration around explicit hypotheses and accumulates evidence across multiple events.

\vspace{.5em}
\noindent{\textbf{Limitations and Future Work}}.
First, our extraction of the Kuroshio jet axis and mesoscale eddies relies on existing feature-detection algorithms and remains sensitive to parameter settings and complex frontal structures. 
Second, our current implementation focuses on selected spatial and temporal subsets rather than scalable processing of the full-resolution simulation, and the computation of composites and statistical evidence requires further optimization. 
Finally, our evaluation is based on a demonstration with two domain experts and therefore provides only preliminary evidence of the system's usefulness. Future work will improve the robustness of feature extraction and characterize its uncertainty, develop parallel and incremental computation for larger datasets, and extend the workflow to longer periods, additional variables, and other frontal regions. We also plan to conduct broader task-based studies with oceanographers to examine how well the system supports the formulation, testing, and refinement of scientific hypotheses.


\bibliographystyle{abbrv-doi}

\bibliography{ref}
\end{document}